# Photo-Ionization Induced Rapid Grain Growth in Novae

Steven N. Shore[1,*] and R. D. Gehrz[2]

[1] Dipartimento di Fisica "Enrico Fermi", Università di Pisa, Pisa I-56127, Italy
e-mail: `shore@df.unipi.it`

[2] Department of Astronomy, University of Minnesota, 116 Church Street SE, Minneapolis, MN 55455 USA
(gehrz@astro.umn.edu)
e-mail: `gehrz@astro.umn.edu`

**Abstract.** An elementary model is developed for grain growth during the expansion stage of nova ejecta. Rather than suppressing grain formation, we argue that agglomeration of atoms by dust nuclei proceeds kinetically through induced dipole reactions in a partially ionized medium. Ionization of a cluster increases once the ejecta become transparent in the ultraviolet which triggers runaway grain growth. This mechanism may also be important in other dust-forming hot stars such as Luminous Blue Variables and Wolf-Rayet systems.

## 1. Introduction

Dust formation is a poorly understood feature of some classical nova outbursts (*e.g.* Gehrz 1988; Evans 1997; Gehrz, Truran, Starrfield, & Williams 1998; Gehrz 2002; Evans 2002). One subclass, the prototype of which is DQ Her 1934, shows a deep minimum in the optical light curve, lasting for months, that usually begins around 60 to 100 days after the rise to visible maximum. This has been interpreted as the rapid formation of large grains, which render the source optically thick until the expansion clears out the environment. The rise in the visual optical depth of the dust shell is remarkably fast, suggesting that grains grow with very high efficiency once nucleation has occurred. The subsequent light curve evolution is related to the combined effects of decreasing column density due to the shell expansion and a decrease in the mean grain size (Gehrz et al. 1980a,b). As the shell becomes optically thin, the light curve returns to the same exponential decay it originally followed. Gallagher & Starrfield (1977) first proposed that this decay occurs at constant bolometric luminosity, as subsequently demonstrated by multiwavelength studies of novae (cf. Shore 2002), and Gehrz et al. (1980a,b) suggested on the basis of the development of the infrared emission during early stages of the outburst that the decrease in grain size is related to the increased intensity of UV flux absorbed by the ejecta at a time when the density of condensibles has been significantly reduced by the shell expansion.

We should note that some of the IR emission seen during outburst has been attributed to environmental cold gas but late time observations in the far infrared of at least one recent dust former, V705 Cas (Nova Cas 1993) with ISO did not find a detectable continuum signature (Salama et al. 1999). Studies of the infrared spectral development are outlined by Evans et al. (1997), Lynch et al., (1997), Gehrz et al. (1998, and references therein). In the few novae that have been well observed before the onset of a dust forming event, molecular absorption from CN has been reported (for instance in DQ Her 1934 and V705 Cas) and infrared CO emission has also been detected in several outbursts (Gehrz, Hackwell, & Briotta 1976, Lynch et al. 1997). A few novae, in particular QV Vul 1987, have displayed signatures of different types of grains including amorphous carbon, SiC, hydrocarbons, and silicates silicates (Gehrz 1988, Gehrz et al. 1998). Thus the precursors for grain formation, molecules, are detectable long before the onset of the main growth stage. However, millimeter wavelength searches for molecular *survivors* of the dusty epoch using millimeter lines of CO have thus far been unsuccessful (Albinson & Evans 1989; Shore & Braine 1992; Weight et al. 1993; Nielbock & Schmidtobreick 2003).

Not all novae form dust. Gallagher (1977) suggested that ionization of the ejecta might suppress grain formation but extensive observations of many recent novae suggest that this ionization may actually *promote* the rapid appearance of large grains. It cannot be a coincidence that the epoch of the dust forming event coincides with another feature of the photometric variations: this is the stage at which the near ultraviolet becomes optically thin. In V705 Cas 1993, this was clearly observed for the first time when, using the IUE satellite, the 1200

- 2000A and 2000 - 3200A regions simultaneously peaked and the dust formation event was observed to start (Shore et al. 1994). The UV photosphere, dominated by absorption lines of iron peak elements, decreased rapidly and uniformly in intensity throughout this spectral window with no change in the lines. In other words, the absorption region in which the dust formed was at a greater distance from the central star and therefore at much lower density than the UV emitting material. It was inferred that the dust event took place when the gas reached the Debye temperature, in this case for silicates since no absorption feature was detected at 2175Å, but the shell was strongly irradiated by ultraviolet at the time and the ionization of the ejecta continued to increase as indicated by the development of the optical emission line spectrum (see Gehrz et al. 1992). Therefore, it appears that the dust formation may have been triggered, rather than suppressed, by, the presence of weakly ionizing radiation. In this note, we suggest a possible formation mechanism for large grains: ionization-mediated kinetic agglomeration of atoms onto molecules and small grains through induced dipole interactions.

## 2. The Role of Photo-ionization and Induced Interactions

One mechanism for atomic accretion by clusters in a partially ionized gas is through induced dipole interactions. A large literature has developed in the last few years related to dusty plasmas, stimulated in part by solar system phenomena and also plasma processes related to fusion and industrial and meteorological applications (*e.g.* Watanabe 1997, Harrison 2000). Hollenstein (2000) discusses how the type of grain formed changes with increasing charge, becoming more irregular and fluffy as the agglomeration proceeds. Photoionization has been only occasionally exploited, as in the original suggestion of grain charging as the mechanism responsible for particle levitation in the Saturn ring system where solar UV flux ionizes and evaporates small ring ice particles, generating small charged grains that couple to the planet's magnetic field, and in cometary comae where the charged grains couple to the interplanetary magnetic field (e.g. Horanyi 1996; Verheest 1999; Mendis 2002). Direct measurements have also recently been made under microgravity conditions with an eye to solar system applications (Tytovich et al. 2003). Notably, Kortshagen and Bhandarakar (1999) find that accretion is strongly suppressed if the number density of nuclei is significantly lower than the positive ion density, regardless of the charging mechanism. In general, however, the effects of ionization on grain formation and growth have not been exploited.

Either participating species may be neutral, with polarizability $\alpha_D$, and the interaction potential is $V = \alpha_D E^2$, where $E$ is the local electric field. For a point source with charge $Ze$, $V = Z^2 e^2 \alpha_D r^{-4}$, where $\alpha_D$ is the polarizability, while for a grain of radius $a$, $E$ is replaced by $Z_g a r^{-3}$. The cross section for a point charge interaction is approximately:

$$\sigma \approx (\frac{Z^2 e^2 \alpha_D}{m_X v^2})^{1/2} \tag{1}$$

where $m_X$ is the mass of the colliding atom with velocity $v$ (for an example of a recent quantum mechanical calculation, see e.g. Xu et al. 2001). This gives a rate for number density $n_X$ that is independent of temperature (except, perhaps, for an excitation energy):

$$R_{gX} = <\sigma v> n_X \approx (\frac{Z^2 e^2 \alpha}{m_X})^{1/2} n_X, \tag{2}$$

where the average is taken over the thermal velocity distribution of the impacting atoms. This is the Langmuir rate that is used in ion chemistry. For a charged grain the rate is weakly dependent on temperature:

$$R_{gX} \sim \alpha_D Z_g^{2/3} a^{2/3} T^{1/6} m_X^{-1/2} n_X. \tag{3}$$

When the gas is partially ionized, the induced rates proceed much faster than those obtained from homogeneous nucleation (once the seeds are formed) since the saturation vapor pressures are very low in the expanding ejecta. We therefore assume that cluster growth proceeds kinetically. If the ionization potential depends on the size of the cluster, in particular if it decreases with increasing size (the most loosely bound electrons become less bound with increasing size) then there is a runaway. The ionization remains constant or increases (the work function for the grain decreases as the grain grows) even as the otherwise neutralizing atoms are added to the cluster so the rate of accretion grows.

Two conditions must be met simultaneously. The grain temperature must be at about, or lower than, the Debye limit for stability and the rate of ionization must be balance the rate at which atoms stick to the surface for growth. Observations confirm that the grain temperatures are ≤1000 K during the rapid growth phase (cf. Gehrz 1999, 2000); since the Debye temperatures for graphite and silicate grains are 750 K and 500-600 K, respectively, this is consistent.[1] The ionization potential for most atomic species of interest, especially carbon, lie in the mid-UV, around 7 eV. For this reason, the gas is mainly neutral in the "iron curtain" interval – during which the molecular seeds can form – and becomes increasingly ionized as the line opacity in the ultraviolet decreases (Beck et al.1995). The grains, on the other hand, have lower work functions, around 4 eV, and can charge even under exposure to relatively soft radiation. The ionization rate for the grains is:

$$R_{pe} = \gamma_{pe} \pi a^2 \int_{\nu_0}^{\infty} Q_{abs}(\nu) F_\nu Y_{pe} \frac{d\nu}{h\nu}$$
$$\approx \frac{1}{4} a^2 f_{UV} f_{ej} \frac{L_{Edd}}{R^2} < \frac{Q_{eff}}{h\nu} > Y_{pe} \tag{4}$$

where $\gamma_{pe}$ is a geometric grain factor and $f_{ej}$ is the filling factor for the dust forming gas, $R$ is the radius of the shell at time $t$, $Q_{eff}$ is the absorption efficiency (which depends on composition), $f_{UV}$ is the fraction of the luminosity above the ionization limit, and $Y_{pe}$ is the photoelectric yield of the grains (see Dwek & Arendt 1992). For $R \approx 10^{15}$cm, typical of the epoch of rapid dust growth, for $Q_{\rm eff} \approx 1$ and $Y_{pe} \approx 0.01$, the rate is $\approx 10^3 f_{UV} f_{ej}$ s$^{-1}$. The grain ionization rate therefore scales as:

$$R_{pe} \approx 10^9 <Y_e Q_{eff} f_{ej} f_{UV}> (\frac{a}{0.1\mu})^2 (\frac{R}{10^{15} cm})^{-2} (\frac{L}{L_{\rm Edd}(1.4 M_\odot)}) \ . \tag{5}$$

---
[1] The referee noted, however, that for V705 Cas the dust temperature was about 1200 K at the time of condensation.

Here we assume that the optical depth at about 4 eV is vanishingly small at the moment of irradiation and the underlying source has a spectrum typical of an expanding nova atmosphere at the end of the iron curtain phase with a color temperature greater than about $1.5 \times 10^4$ K.

The rate of atom-grain collisions, assuming induced dipoles and charged grains, is:

$$R_{gC} \approx 1.4 \times 10^{-5} S_{gC} \alpha_0^{1/2} n_C (\frac{a}{0.1\mu})^{2/3} Z_g^{2/3} (\frac{T}{10^4 K})^{1/6} \tag{6}$$

where $S_{gC}$ is the sticking coefficient for the carbon atom on a grain site and $\alpha_0$ is the scaled value of the polarizability (units of $10^{-24}$Å$^3$). At $\approx 100$ days, the radius of a typical nova shell is $10^{15}$cm for a maximum ejection velocity of about 3000 km s$^{-1}$; these values are upper limits for the dense material in CO novae – for which the velocities are often as low as 1000 km s$^{-1}$ and the dust growth begins at closer to 60 days – but they provide a useful upper bound for the required atomic density.

The limiting condition for grain growth is that the charge should not increase beyond the Rayleigh limit at which the electrostatic self energy of the grain exceeds the binding, thus the charge should remain roughly constant with each accreted atom. Imposing the condition that $R_{pe} \approx R_{gC}$ gives $n_c \approx 10^{10} f_{ej} \alpha_0^{-1/2}$ cm$^{-3}$. At the epoch of dust growth, the hydrogen density has about the same value so the required $\epsilon_C = n_C/n_H \approx 10^3 - 10^4 \epsilon_{C,\odot}$. Thus, for a runaway to occur, the carbon abundance must be substantially enhanced over solar values. This is consistent with abundance determinations for CO novae, for which the observed $\epsilon_C/\epsilon_{C,\odot}$ can be as large as $10^4$ (e.g. Gehrz 2002).

It is also possible that the grains, rather than the atoms, may develop induced dipoles. Polarization measurements of grains yield far larger $\alpha_D$ than for atoms (Broyer et al. 2002). For instance, for Si$_n$, up to 50 particle clusters, $\alpha_D \approx 4$Å$^3$ but for fullerenes that may be precursors of graphite grains – particularly C$_{60}$ and C$_{70}$, measurements yield $\alpha_D$ of 76.5±8 and 102±14Å$^3$, respectively. This provides an additional channel for agglomeration reactions at the initial stages of the expansion with neutral grains interacting with carbon ions. Metallic fullerenes have up to an order of magnitude higher $\alpha_D$. Thus the initial stages of the runaway grain growth may be due, in part, to the small fraction of the atomic ejecta that is ionized. However, since the work functions for clusters are typically smaller than the ionization energies of individual atoms, this is a very inefficient channel since the number of reacting particles will be much lower than during the later stage when atomic neutrals react with charged grains. Indeed, this may be the limiting factor for grain growth, as originally conjectured: when both the grains and the atomic gas are mostly ionized reactants must overcome a coulomb barrier and this is not possible at these temperatures and densities.

We now consider a simple model for the grain growth, in fact a very general one given the cross sections we have derived above. In outline, for a dense (nonfractal) grain, the mass scales as $a^3$ so the continuity equation for dust growth when accreting thermally impacting particles is $\dot{m}_g \sim \sigma_{gC} m_C n_C v_T$ where $\sigma_{gC} \sim a^{2/3}$ and $v_T$ is the thermal velocity for the ejecta. Then substituting $\dot{m}_g \sim a^2 \dot{a}$ and assuming the usual linear velocity law for the ejecta so that $n_C \sim t^{-3}$, for $Z_g$ = constant we find:

$$a \sim a_0 (1 - (\frac{t_0}{t})^2)^{3/5}.$$

Here $t_0$ is the initial epoch for grain formation and $a_0$ is the asymptotic grain radius. Since the optical depth varies as

$$\tau_g \sim n_g a^2 R = \tau_0 (1 - (t_0/t)^2)^{6/5} (t_0/t)^2, \tag{7}$$

the maximum optical depth, $\tau_0$, occurs at $t_\star = \sqrt{5} t_0$ (we remark that if the charge depends on the size of the grain as $Z_g \sim a^n$, this epoch is $t_\star = [(2n+11)/(2n+5)]^{1/2} t_0$ for constant $T$. There is some evidence that $n > 0$ for laboratory studies (see Maisels, Jordan, & Fissan (2003) and Mendis (2002) for solar system dust). The weak temperature dependence of the collision cross section insures that this is a small effect and can be neglected and, actually, this is a generic result. Observations show grain growth beginning at around 60 to 100 days, so we expect the maximum optical depth to occur at around 140 to 240 days.[2] To show an application of our model, in Fig. 1 we display the visual data for V705 Cas during the 1993-1994 dust forming episode. Figure 2 shows the results of our model scaled to the time of dust growth ($\approx 60^d$) and the magnitude at that time. The only additional free parameter, $\tau_0$, was chosen to approximately match the depth of the event.[3]

## 3. Discussion

Dust formation was first considered in novae by Clayton and Hoyle (1976) who used homogeneous nucleation theory (Donn et al. 1968) to predict the supersaturation for grain growth. Their conclusion, that to form dust requires a significant carbon enhancement, has remained valid. Gallagher (1977) argued that grain growth is stopped by the increasing ionization of the ejecta, a point echoed in the model by Kortshagen and Bhandarkar (1999). Joiner (1999, see also Joiner & Leung 1999) modeled grain formation as an essentially chemical process in which growth of embryonic clusters competes with photodissociation. Clayton, Liu, & Dalgarno (1999) noted that dust growth can be accelerated by radioactive species in supernova ejecta that ionize the gas and promote cluster formation. This is analogous to the well known phenomenon of condensation clouds appearing immediately after the burst of ionizing radiation released in terrestrial nuclear explosions (see, e.g. Gensdarmes et al. 2001).

---

[2] Electrons play a role here in charging the grains as well (see e.g. Piel & Melzer 2002). Hot electrons can tunnel through small grains, effectively ionizing them, while the energy loss in larger grains reduces the electron yield. See the review by Mendis (2002). In the solar system, thermionic emission is also invoked, and this process may occur for small nuclei during the initial stages of the outburst when the UV is strongly suppressed.

[3] We don't want to make too much of this result in detail. The derived mass implied by scaling is about $10^{-7}$ to $10^{-6} M_\odot$ for the dust formed in this event. This is of the same order as the mass derived by Gehrz et al. (1995) but about ten times the result reported by Evans et al. (1997). However, considering the very schematic comparison, we suggest that even this rough agreement is interesting.

The results for the ionization structure of nova shells obtained by Beck et al. (1995) are particularly germane to our discussion (see also Beck 1993). They find, using fully NLTE line blanketed moving atmosphere models, that the carbon ionization is not complete during the first few months of the outburst. In general, this is the stage when the UV is opaque, although for fast CO novae the optical depth decreases rapidly, and suggest that ion reactions may produce complex chemical products. The models use spherically symmetric, homogeneous shells. In most scenarios, however, grain growth is assumed to occur in knots that provided environments shielded from the ultraviolet from the central source as well as higher densities that enhance the reaction rates. The shadowing has also been invoked to account for the strength and persistence of the O I 1300 resonance lines late into the outburst (*e.g.* Williams 1992). Indeed, such structures are frequently observed. Nova shells are complex, as seen from both the spectra obtained during the early stages of outburst and in later, spatially resolved images. But in general the filling factors for the dense gas are small, of order a few percent, and the density enhancements are not more than a factor of about ten (see Vanlandingham 1998; Schwarz 2000).

It is possible that the same mechanism can operate in hot stellar winds, especially the Luminous Blue Variables (LBVs). During outbursts, these stars display many of the characteristics of novae during the optically thick stage: a pseudophotosphere forms that drops the effective temperature from that of a late O or early B supergiant to about $1.5 \times 10^4$K due to the saturation of the iron peak lines. The increased density from the enhanced wind coupled with the small residual ionization may suffice to initiate the same grain growth seen in novae. A discussion of the details requires precise radiative transfer calculations in the winds but the situation is similar. For WR stars, as discussed by Williams et al. (1987), it is possible that the same mechanism operates during outburst when the ionizing radiation is reduced in the range of the carbon continuum.

*Acknowledgements.* We thank the referee, Nye Evans, for a very careful and helpful reading, and Jason Aufdenberg, Paola Caselli, Peter Hauschildt, Eric Herbst,, Ted LaRosa, Dave Lynch, Greg Schwarz, Sumner Starrfield, and Karen Vanlandingham for valuable discussions and correspondence. Some of these ideas were developed during the Sitges meeting (2002) and we warmly thank Margarita Hernanz and Jordi José for arranging such a fruitful and exciting conference. SNS thanks William Feighery, IUSB Chemistry department, for many valuable discussions. thank the Dr. Janet Mattei and the AAVSO for supplying the data for V705 Cas, and the AFOEV use of their archive. The community truly benefits from such superb public data archives. SNS acknowledges support from NASA and a research award from Indiana University; RDG acknowledges support from the NSF, NASA, and the US Air Force.

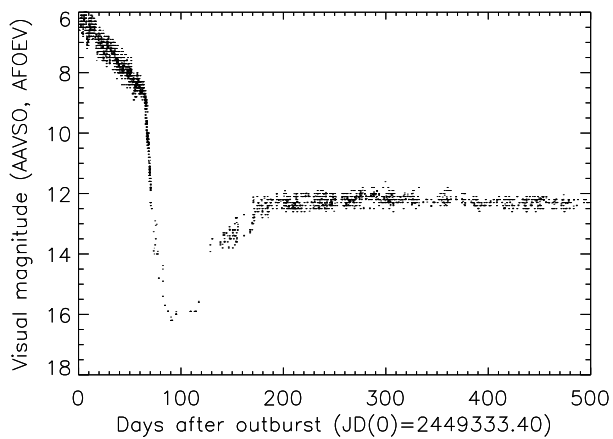 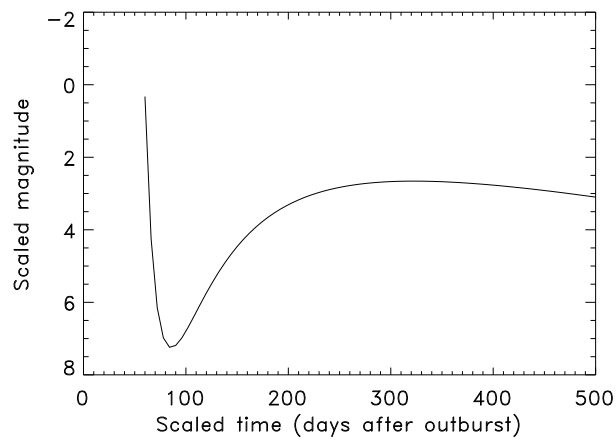

**Fig. 1.** Visual light curve for V705 Cas based on AAVSO and AFOEV data.

**Fig. 2.** Simulated light curve for parameters approximating those of V705 Cas using eq. (7). The maximum scaled optical depth was 25, a simple exponential light curve was assumed ($t_{\rm decay} = 200^d$), and the onset of dust formation was set at 60 days. Magnitudes are differential realtive to the time of the dust formation (V≈9 mag).